# Tunable polarization components and electric field induced crystallization in polyvinylidenefluoride (PVDF); a piezo polymer

Ronit Ganguly[2], Vijayabhaskar Gunasekaran[2], Kumaraswamy Miriyala[1], Soumya Bandyopadhyay[1], Saswata Bhattacharya[1], Amit Acharyya[2] and Ranjith Ramadurai[*1]

1. *Department of Materials Science and Metallurgical Engineering, Indian Institute of Technology Hyderabad, Kandi, Sangareddy, Telangana, India – 502285*
2. *Department of Electrical Engineering, Indian Institute of Technology Hyderabad, Kandi, Sangareddy, Telangana, India – 502285*

Polyvinylidenefluoride (PVDF) a semicrystalline pieozoelectric polymer was synthesized with varying process conditions and its ferroelectric domain orientations were studied using piezoresponse force microscope (PFM). PVDF thin films fabricated using tape casting technique with precursor solutions of varying viscosities reveal that the polarization components transform from a dominant planar component to an out-of-plane polarization components with increase in viscosity. Interestingly the planar components possessed a head to head or tail to tail kind of paired domains separated by a distance of ~ 380-400nm. The electrostatic energies computed by numerically solving the electrostatic equilibrium equation for the electrically inhomogeneous system are in good correlation with the experiments. On increment of electric field, the domains were observed to grow in size and shape which indicates amorphous to crystalline transformation in the case of PVDF. Such transformation was evident from x-ray diffraction studies performed in-situ in the presence of an applied electric field.

---

[*] Corresponding author email: ranjith@iith.ac.in

Piezoelectric polymers are gaining attention as smart materials in various applications, such as sensors, actuators, energy harvesters and biomedical devices. Amid the few, poly(vinylidenefluoride) (PVDF) and its copolymers are one of most widely used piezoelectric polymer. This semi-crystalline polymer shows a complex structure and there have been ample amount of studies on PVDF and its copolymers, focussing the piezoelectric property.[1,2] PVDF is known to stabilize in five distinct crystalline phases related to different chain conformations α, β, γ, δ and ε, in which the most investigated phases are α, β, γ.[1,3,4,5] β–phase is known to exhibit superior piezoelectric properties due to its non-centrosymmetric crystalline state with the dipole moment of two chains containing C-F and C-H in the unit cell adding up resulting in a net dipole moment perpendicular to the carbon backbone with fluorine as negative and hydrogen as positive poles.[2,6,7,8]

There are various techniques to fabricate PVDF thin films and its properties are known to be sensitive to the methods adapted.[1,9] Literature suggests that the transformation occurs in a given sequence α → γ → β with varying temperature and/or strain induced processing.[5,10,11,12] Major characteristics like, morphology, optical transparency and mechanical properties of PVDF films synthesized through solution casting technique is known to be dependent on the viscosity of initial solution.[13,14,15]

Microscopic studies reveals that the melt crystallized PVDF polymer film consist of spherulites, which are basically stacks of lamella with thickness 10-20nm and possess a non-crystalline (liquid like) region between the crystalline lamellae.[16] Such structural heterogeneity gives rise to a dielectric amorphous region between the homogenous polar regions. The ferroelectric property of these semi-crystalline polymers are known to arise from the crystalline polar regions having lamellar structure. However to enhance the ferroelectric/piezoelectric properties, completely crystalline PVDF-TrFE co-polymers were studied. There have been limited studies detailing the presence of nanoscale ferroelectric domains and switching in PVDF-TrFE (Triflouroethylene) films by piezoresponse force microscopy (PFM).[17] The studies involving both the of out-of-plane and in-plane polarization components suggests the presence of nano-mesas and the polarization switching occurring around the back bone of the polymer chain.[17] In addition, polarization contrast with randomly shaped regions separated by un-polarized boundary regions were observed and the effect of poling on the polarization switching has been reported on the copolymer.[18] External electric field has a significant influence on the structural transitions in PVDF and its copolymers.[19,20,21] Electric poling induced crystallinity has been studied earlier through x-ray diffraction studies and confirmed an amorphous to crystalline transitions in PVDF-TrFE.[22] The presence of domain patterns with uniformly polarized regions were further confirmed by polarized Raman and Infrared spectra.[23,24] In this paper we address the tunability of polarization components offered by the preparation conditions of semi-crystalline PVDF film. The films with dominant planar polarization components are utilized to understand the domain features and the transformation of domains under applied electric field. The experimental observations of domain separation distance agree well with the numerical solutions of Poisson equation for electrically inhomogeneous medium. We also demonstrate the enhancement of crystallinity in the presence of electric field by performing x-ray diffraction studies in the presence of the electric field.

Commercially available PVDF (Sigma Aldrich) powders and N-Methyl-2-pyrrolidone (NMP) (Sigma Aldrich) solvent with appropriate weight proportion were mixed in a magnetic stirrer to obtain the solution of PVDF. The PVDF solutions with varying viscosities were prepared by varying the weight percentage of PVDF powder. The viscosity of the PVDF solutions were studied using a Rheometer (Physica MCR 301) at shear rate of 100/s. PVDF solution with varying viscosity range 2 to 8 Pa.S was synthesized and further utilized for the fabrication of thin films. The solutions were casted to films on Aluminium foil/glass slide by the conventional doctor Blade technique and further baked at $60^0C$ for 8 hrs. In order to obtain β phase the films were annealed at $90^0C$ in an oven and later the films used for further studies.

The inset of Figure 1 shows the X-ray diffraction pattern of a PVDF thin film synthesized from the precursor solution of 4.6 Pa.S. The peak at 2θ = 20·2 which is relative to the sum of diffraction from the planes 200 and 110 confirms the presence of β-phase in PVDF films and the shoulder peak at 18·5 corresponding to 020 plane which indicate the existence of small percentage of γ-phase.[1,4,10,25,26,27] The films prepared from varying viscosities seem to have no change in the percentage of mixed phases present, however a minor enhancement in the crystallinity of PVDF films was observed when synthesized from higher viscous solutions.

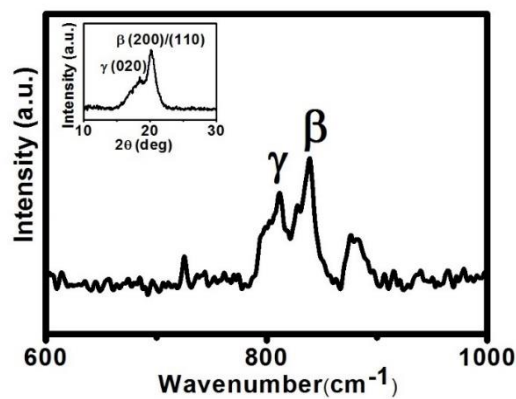

FIG. 1. XRD graph and Raman spectra.

Fig. 1 shows the Raman spectra of pure PVDF thin film. The Raman spectra is dominated by a band at 839cm$^{-1}$. The high intensity peak at 839 cm$^{-1}$ confirms the high percentage β-phase for annealed films. The Raman bands at 811 cm$^{-1}$ in PVDF films, corresponds to γ phase and the peak at 882 cm-1 corresponds to the mixed phases. [19,28] It is known that the solution casting method often gives rise to mixed phases which is also evident from our results. Since our interest is mainly on understanding the ferroelectric domains of pure PVDF with its semi-crystalline nature we continued our studies with PVDF and not sticking to the methods known to enhance the β–phase.[1,27,29,30,31,32,33] From above characterization we can conclude that the PVDF films in consideration mainly consists of β-phase along with minor fraction of other γ phase. Fig. S1 of the supplementary document shows the variation of piezoresponse amplitudes with varying viscosity. It was observed that in our PVDF films throughout the viscosity range the in-plane domains are dominant but as the viscosity increases there is significant rise in the out-of-plane domains. This can be attributed to increase in density along with viscosity. Increase in density causes the reorientation of chain structure of PVDF polymer which in turn leads to change in domain orientation, thus transforming 2D to 3D domain orientation. Further increase in density results in equal contribution of both in-plane and out-of-plane domains.

In addition, PFM studies reveal the presence of the orientation of various polarization components in which, the piezoresponse amplitude and phase image of the respective films gives better understanding of the ferroelectric domain patterns present. Fig. 2 shows the PFM images of PVDF film prepared from the precursor solution with a viscosity of 4.5 Pa.s. The various features like morphology, out-of-plane (OP) piezoresponse and phase, in-plane (IP) piezoresponse and phase are presented in Fig. 2. Interestingly the Fig. 2(e) reveals the presence of dominant IP polarization components in this particular film. It evidently shows the presence of paired regions with opposite domain orientations either left or right with respect to the probe tip. The domain are out of phase to each other by 180$^0$ and are separated by a distance of 380-400nm. This distance observed is relatively larger in comparison with conventional ferroelectric domain separations. However, conventionally the ferroelectric domain walls are 180$^O$ with either head-to-tail or tail-to-head configuration. In this case we believe that the intermediate region presumably amorphous is capable of storing excess charge,[34,35,36] and hence maintaining a distance of few 100 nm to stabilize the head to head or tail to tail configuration. Presence of such paired domains were distributed uniformly throughout the sample and the scalable nature of the domain patterns was evident from the large area scans and presented in supplementary Fig. S2.

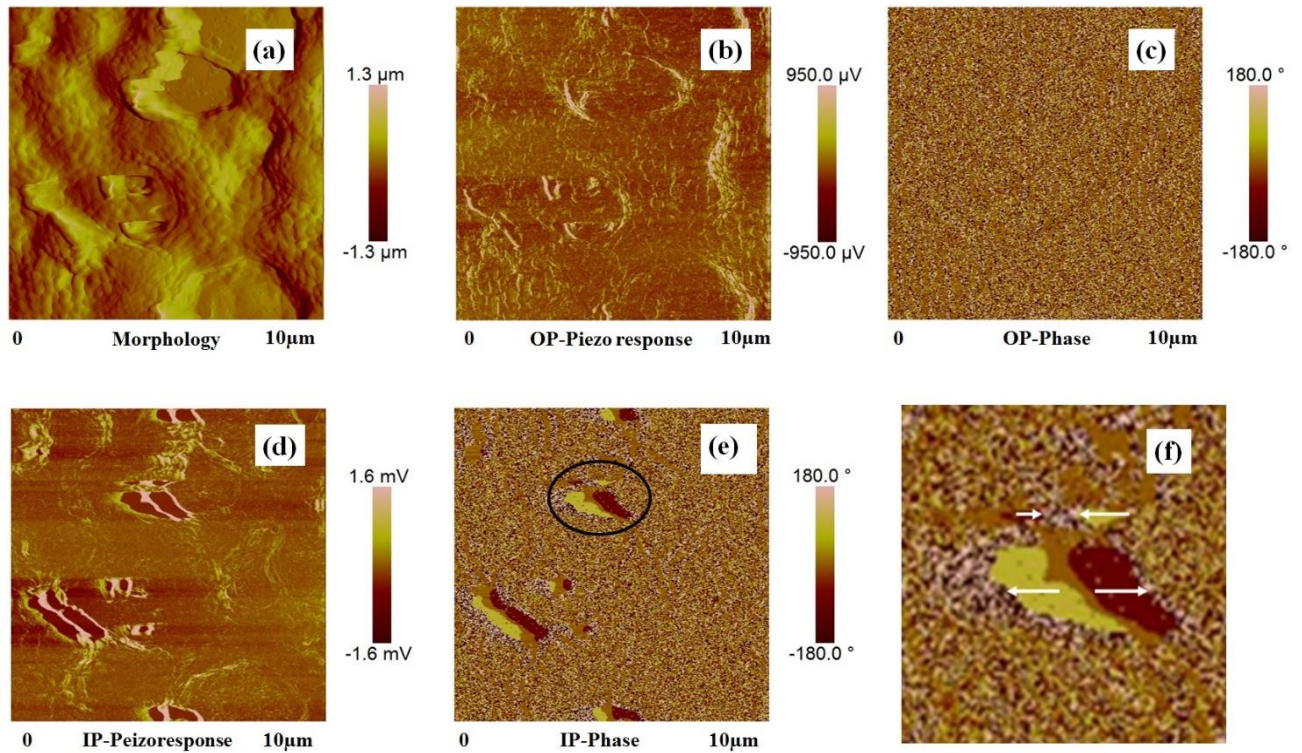

FIG. 2. PFM images of PVDF films (a) Morphology (b)Out of Plane (OP) Piezo response (c)Out of Plane phase (d)In-Plane (IP) Piezo response (e) In-Plane Phase (f) zoomed region of 2(e) showing the nucleation of opposite orientations adjacent to the larger domains

As these planar domains are of similar to same charges facing each other, it results in an electrostatic interaction within the pair of domains. Hence, electrostatic force from the domains are expected to influence the nucleation of other domain in the surrounding area. It is evident from Fig. 2(f) that a criss-cross configuration of domain orientation was observed between the adjacent pair of domains. The arrow marks are representative guide for the eyes and an opposite representation with 180º rotation is equally plausible.

To understand the electrostatic interactions between the polar domains separated by a non-polar medium and the equilibrium separation distances between the domains we constructed checkerboard-type domain configurations similar to those in PFM images and implemented an inhomogeneous Poisson solver to compute electric field and electrostatic energy density. Fig.

3(a) shows one typical initial configuration of in-plane ferroelectric domains separated by a paraelectric matrix (red). Each pair of domains (denoted by yellow and black) in the checkerboard structure shares a 180° domain wall. The separation distance along x-axis is kept constant while that along y-axis is varied systematically using a step size of 10 nm. We compute the electrostatic energy of each configuration to study the stability of the configurations as a function of vertical separation distance. The minimum in electrostatic energy is used as a criterion to determine the most stable configuration. We used an average domain width of 600 nm and a horizontal separation distance of ~250 nm to mimic experimentally observed domain patterns.

Electrostatic energy of a given configuration is obtained by solving the electrostatic equilibrium equation for electrically inhomogeneous systems[37]:

$$\nabla \cdot [\epsilon(\bm{r})\nabla\phi(\bm{r})] = \nabla \cdot \bm{P}(\bm{r}), \tag{1}$$

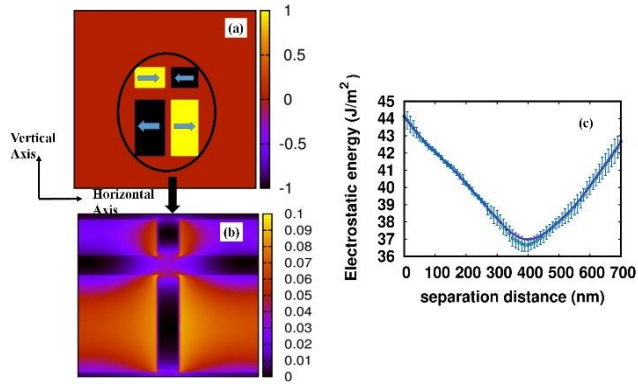

FIG. 3. (a) A typical configuration of the ferroelectric domains to compute electrostatic interactions between domains (similar to the domain arrangement observed in experiments). (b) Electrostatic energy distribution when the vertical separation distance between the domains is 390 nm (c) Electrostatic energy vs vertical separation distance (along y-axis).

where $\epsilon(\bm{r})$ is position-dependent dielectric permittivity, $\phi(\bm{r})$ is the electric potential distribution and $\bm{P}(\bm{r})$ denotes the inhomogeneous polarization field[37]. The electric field is given as $\bm{E} = -\nabla\phi$ and the electrostatic energy is given as

$$F_{electric} = \frac{1}{2}\epsilon(\bm{r})E^2 + \bm{P} \cdot \bm{E}. \tag{2}$$

The electrostatic equilibrium equation was solved using INTEL MKL Library assuming periodic boundary conditions in absence of external field.[38] Our calculations show that the electrostatic energy of a configuration is minimum when the vertical separation distance varies between 350 – 450nm. Further increasing the distance increases the energy of the system (Fig. 3(b)). A close agreement between the experimental findings and numerical results indicate that the minimization of electrostatic interactions is the primary reason behind the arrangement of in-plane polar domains separated by an amorphous layer. Such a configuration stabilizes because the amorphous layer (of 350-450 nm width) can effectively compensate the excess charge associated with the head-to-head / tail-to-tail configurations of the in-plane domains when the separation distance is ~390nm.) The computationally obtained result almost matches with the minimum separation distance of ~400nm between the pair of domains that is obtained from the experimental findings.

The β phase of PVDF is thermodynamically meta-stable phase with all trans (TTTT) zigzag chain conformation.[7,27] PVDF exhibits ferroelectric property as its electric dipole due to C-F and C-H can be changed in the presence of external electric field,[17] which is confirmed by the ferroelectric hysteresis loop. The SS (Switching spectroscopy)-PFM studies performed in all the samples at various locations confirm the presence of ferroelectric hysteresis with switching characteristics.

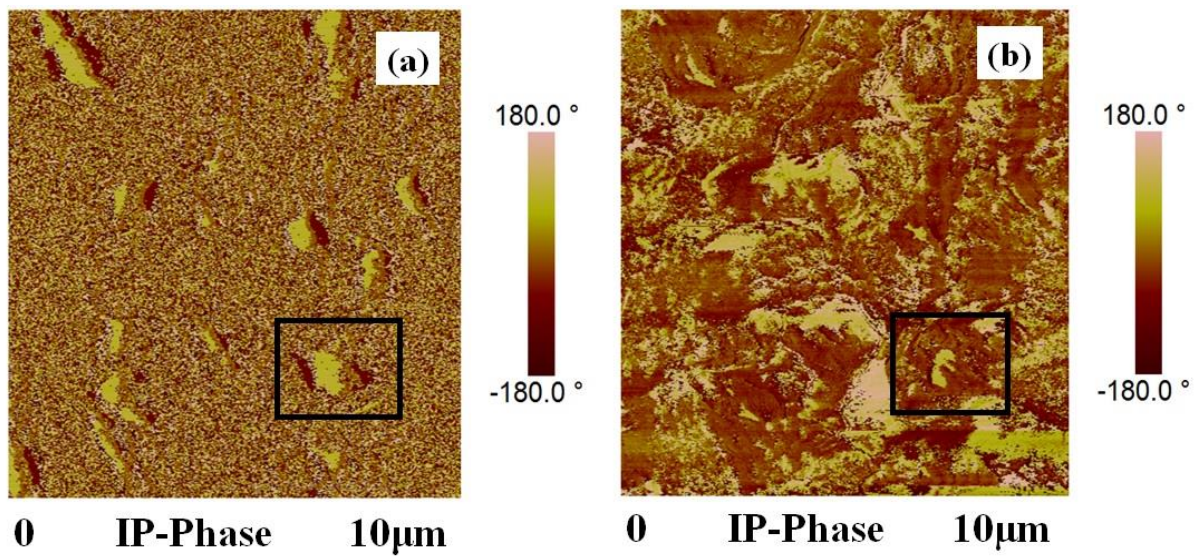

FIG. 4. The PFM in-plane response induced under Tip bias for (a) No Bias (b) 5V DC

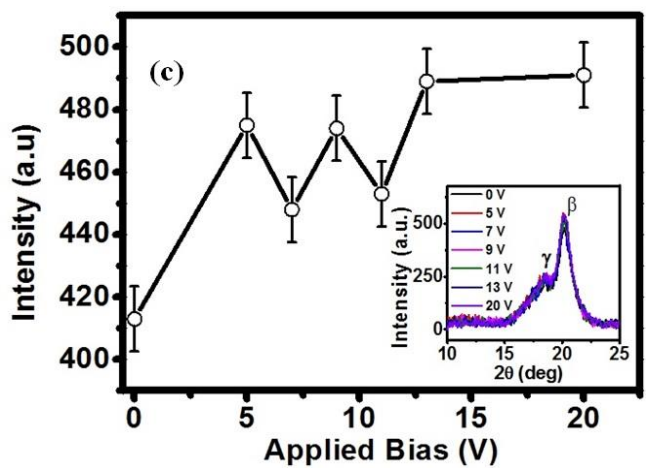

FIG. 5. XRD peak Intensity variation in presence of DC bias.

Earlier studies suggest that application of electric field during the preparation of PVDF samples had its effect on the degree of crystallinity.[39] It is known that the external electric field contributes in direction perpendicular to carbon backbone causing dipoles to rotate only with in-plane perpendicular to backbone.[33] In this study on application of dc-bias across the PVDF thin films a variation in the domain pattern was observed which was similar to the domain patterns observed in completely crystalline PVDF-TrFE co-polymers.[24] Fig. 4(a) and (b) shows the PFM phase image with and without bias. During PFM scan when the unpoled PVDF is subjected to sample bias 5V and then further to 10V, we observe there is a significant change and uniform distribution in the domain contrast which also indicates enhancement in the crystallinity of the film. The enhancement was prominent in the in-plane compared to out-of-plane polarization. The electrostatic force within the PVDF during PFM alter the polymer chain causing monomer to rotate in the direction of electric field thus leading to change in the arrangement of irregular amorphous region and converting them to a crystalline region. Since Pure PVDF is expected to be semi-crystalline such a transformation of IP-phase contrast in the domains is expected to be associated with an amorphous to crystalline transformation. In order to confirm such a transformation if any present, we performed the XRD studies in the presence of an applied electric field.

PVDF samples with thickness of ~20μm on glass substrate with both platinum electrodes on top facilitating an applied field in planar configuration was fabricated. Sample was aligned carefully such that the x-ray scans only the region between the electrodes. XRD on the sample was performed at 0V - 20V and the corresponding change was observed as a function of the peak intensity. In Fig. 5 as voltage increases from 0V to 5V a significant rise in β phase peak intensity was observed. The rise was almost 15% when compared to the intensity obtained at 0V. The intensity oscillates around the same value between 7V to 11V and then a further increase in peak intensity is observed up to 20V and saturates further. Previous studies show similar enhancement in intensity when PVDF films were subjected to poling and later analyzed by x-ray diffraction studies.[7] In the present case we have studied the variation of intensity with an in-situ application of external bias and the variation of intensities are in good correlation with earlier studies. This increase in β phase intensity of our PVDF films confirms the increase in degree of crystallinity under application of electric field. The γ peak also follows the similar trend indicating the ratio of mixed phase remains unchanged across the transformation. Hence, the changes observed in the crystal structure due to electric field do not seem to affect the crystal phase of PVDF.

In summary, we studied the polarization behaviour of PVDF film over a viscosity range. The PVDF films exhibits a dominant planar domains with a unique pairing of domains. The amorphous region separating the domains is capable of holding excess charge and hence facilitates head-to-head or tail-to-tail kind of domain pair configuration in PVDF films. The electrostatic energy of such domain configuration was calculated and found to be in good correlation with the experimental observations. The influence of external bias leads to a change in domain configuration and size and is associated with an amorphous to crystalline transformation. Such a transformation was evidently observed in XRD studies in the presence of applied electric field.